# Hybrid plasmonic nanosystem with controlled position of nanoemitters

**Aurélie Broussier[1], Ali Issa[1], Loïc O. Le Cunff[1], Tien Hoa Nguyen[2], Dinh Xuan Quyen[2], Sylvain Blaize[1], Jérôme Plain[1], Safi Jradi[1], Christophe Couteau[1] and Renaud Bachelot[1*]**

[1] *Institut Charles Delaunay, CNRS. Light, nanomaterials, nanotechnologies (L2n) Laboratory. Université de Technologie de Troyes, 12 rue Marie Curie, 10004 Troyes Cedex, France*

[2] CNRS-International-NTU-Thales Research Alliance (CINTRA), 50 Nanyang Drive, Singapore 637553 Singapore

*Corresponding author: renaud.bachelot@utt.fr

Abstract:

Quantum dots optically excited in close proximity to a silver nanowire can launch nanowire surface plasmons. The challenge related to this promising hybrid system is to control the position of nanoemitters on the nanowire. We report on the use of two-photon photopolymerization process to strategically position quantum dots on nanowires at controlled sites. A parametric study of the distance between the quantum dots and the nanowire extremity shows that precise control of the position of the launching sites enables control of light intensity at the wire end, through surface plasmon propagation.

Nanoplasmonics is an area of increasing interest. One of the most recent and promising research topic is hybrid nanoplasmonics that attempts to control the energy transfer between nano-emitters (NE) and surface plasmons (SP).[1,2] SPs are collective electronic coherent oscillations coupled to electromagnetic surface waves that are evanescently confined at the interface between a metal and a dielectric medium. Silver nanowires (Ag-NWs) are metal nanostructure (MNS) which support propagative SPs that can be launched or detected at the nanowire's extremities.[3,4] On the other hand, semiconducting quantum dots (QDs) are well-known as good NE[5] and their optical properties make them suitable candidates for light-emitting hybrid plasmonic nanosystems.[6] In particular, they permit easy design of specific emission wavelengths through their size and composition.[7] The position of the QDs, however, has to be accurately controlled relative to the MNS in order to obtain an efficient and reliable NE-MNS coupling and energy transfer.

Over the past decade, energy transfer between QDs and plasmonic structures (including metal NWs) has been giving rise to many reported studies (e. g.[8–10]). Most of them used spin coating as a deposition technique of QDs onto the plasmonic system.[10,11] Spin coating is a quick and simple solution but does not permit any control of the position of the QDs relative to the MNS. As a result, many samples must be made before obtaining a satisfactory one where QD location is suitable for physical studies of interest. Some other reported studies still used spin coating to deposit the QDs but then used either electron beam lithography[12] or atomic force microscopy[13] to place the nanoparticles at strategic locations. Other articles reported on the use a layer of polyethylene, which captures QDs that are in aqueous solution,[14,15] while others use a ligand molecule to graft QDs onto the metallic nanoparticles.[16–18] Similarly, QDs functionalization can also be used to have QDs on plasmonic structures.[19] Finally, photopolymerization around metallic nanoparticles through the excitation of localized SPs and involved evanescent waves turned out to be an efficient solution to strategically place QDs around them[20]. Following on from this local photopolymerization idea, we propose an even simpler and faster solution for controlling the position of QDs on plasmonic structures that supports propagating surface plasmons. Our approach of NE positioning is based on far-field two-photon photopolymerization (2-PP) of the QD-containing photosensitive materials that was used and described in references. [20,21] In ref. 20, using such a hybrid photosensitive material, we demonstrated local trapping of QDs around MNS, based on the use of MNS evanescent plasmonic near-fields to trigger 2-PP, resulting in a polarization sensitive hybrid nano emitters. In this letter, we report on a new controlled hybrid plasmonic nanoemitter (see illustration in Fig. 1) based on coupling between CdSe/ZnS QDs and propagating SPs that are supported by silver nanowires (Ag-NWs considered as SPs resonators[22]) and observed though their scattering at the nanowire ends.[10,23]

While QDs/Ag-NW coupling has already been reported (e. g.[24]), it is emphasized that, for the first time, we placed the QDs at different distances from the Ag-NW extremity by 2-PP with a spatial precision that allows for well-defined plasmon launching sites and scattered photoluminescence (PL) intensity at the NWs extremity, as well as quantitative determination of the SPs propagation length.

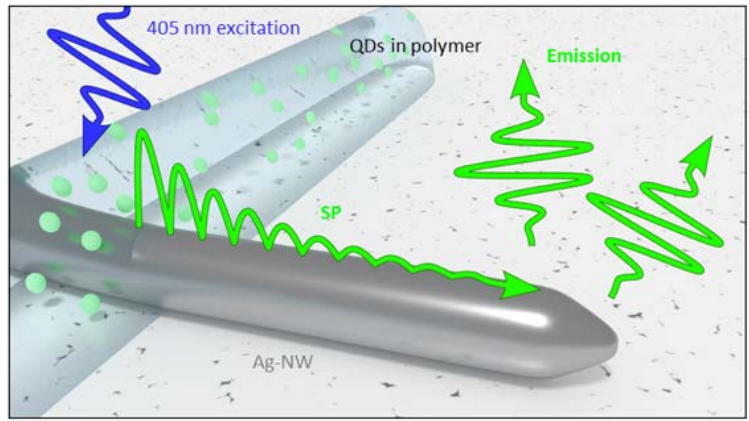

FIG. 1. Representation of the developed hybrid nanosystem based on coupling between light emitting Quantum Dots (QDs) in polymer and Surface Plasmons supported by silver nanowire (Ag-NW).

The Ag-NWs (Sigma-Aldrich, ref 739448) were diluted in an isopropanol solution and have a length dispersion ranging from 5 to 50 µm. Samples were prepared by spin-coating a solution of silver nanowires in isopropanol onto a clean glass substrate at 3000 r.p.m for 30 seconds. Scanning Electron Microscopy (SEM) showed that the average nanowire has a diameter in the 130-160 nm range and is monocrystalline in nature:[25] chemically grown monocrystalline Ag-NWs are characterized by smooth defect-free surfaces and faceted ends, as shown on the SEM images in Fig. 2 (a).

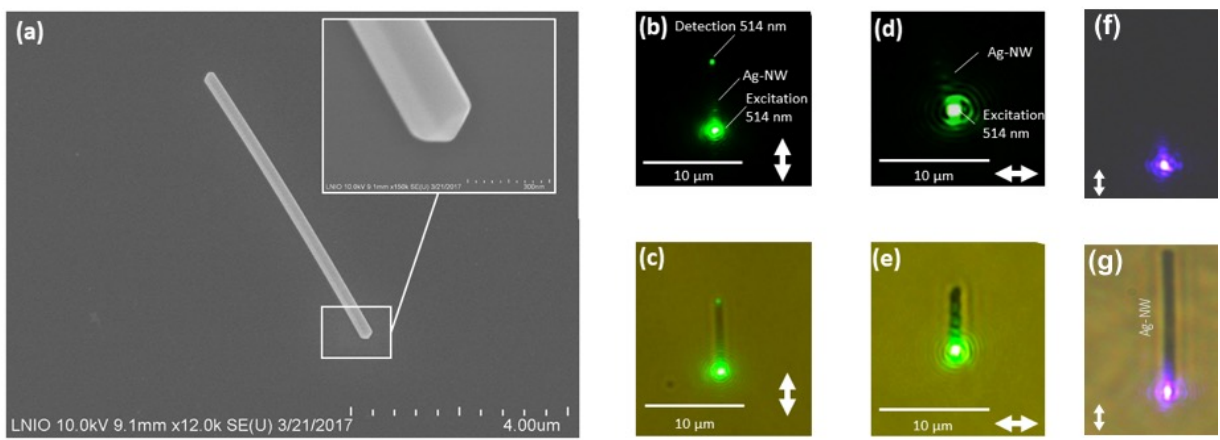

FIG. 2. Used silver nanowires. (a) SEM image of an Ag-NW on Si substrate. Inset: zoom in at the NW extremity. (b) 7µm long Ag-NW illuminated by a focused green beam polarized along the wire axis, (c) same as (b) with white light. (d) same Ag-NW

as (a) illuminated by a focused green beam polarized perpendicular to the wire axis, (e) same as (d) with white light. (f) 5 μm long Ag-NW illuminated by a focused blue beam polarized parallel to the wire axis, (g) same as (f) with white light. The white arrows represent incident polarization direction.

By focusing laser light (λ=514 nm) onto the Ag-NW end, SPs are excited and launched, as evidenced by the green light emission spot at the other end, resulting from SPs perturbation and scattering (Fig. 2 (b-c)). The nanowire end geometry allows one to couple SPs into propagating waves. No SP excitation was observed when the laser is focused on the NW body, far from its extremities, showing that the sharp NW extremities shown in the inset of Fig. 2 (a) enables, through scattering, generation of a large range of high in-plane wave vectors that match the SPs dispersion function at this wavelength. The excitation of SPs modes using this end-scattering coupling scheme is dependent on the polarization of the incident light.[26] Because of their partly longitudinal nature as a wave, SPs can be excited for an incident polarization parallel to the NW axis (Fig. 2 (b,c)), but not for an incident polarization perpendicular to the axis (Fig. 2 (d,e)), confirming that those Ag-NWs are interesting candidates for supporting SPs propagating along their axis in the green.

It should be pointed out that no SPs propagation was observed in case of NW excitation at λ=405 nm which is the wavelength used for QDs excitation in the following (see Fig. 2 (f,g)).

NEs were integrated by local 2-PP of photosensitive formulation made of 1 % Irgacure 819 (IRG 819) and 99 % QDs-grafted pentaerythriol triacrylate (PETIA).[20,21] CdSe/ZnS QDs with an emission wavelength centered on 510 nm were used. The photo-polymerization process was achieved thanks to a femtosecond laser at λ=780 nm focused by a 100x/1.3NA oil immersion objective. This wavelength matches the IRG819 absorption peak at a wavelength of 368 nm. Two photon absorption leads to polymerization reaction when the exposure energy dose exceeds a given threshold $D_{th}$ that is precisely assessed by far-field pre-studied before the experiment.[27]

Since QDs were grafted onto the monomers in solution, they can get firmly trapped within the resulting polymer structure. The remaining liquid unpolymerized QD-containing solution was removed by rinsing with acetone for 10 min, followed by HCl acid for 10 min and isopropanol for 10 min.

Practically, the Ag NWs were deposited on a glass substrate by spin coating. A drop of photopolymerizable formulation was deposited on the sample. Then, a line of QDs-containing polymer was drawn perpendicularly to the NW axis through 2-PP laser scanning (*Nanoscribe GmbH*) with an incident dose slightly higher than $D_{th}$. Fig. 3 shows typical resulting hybrid systems consisting of a Ag nanowire surrounded by written QD-containing polymer lines. These lines have a typical width in the 250-300 nm range but 70 nm wide lines could be obtained under specific conditions at λ=780 nm exposure.[21] Several polymer lines were made on several nanowires in order to carry out a parametric study of the effect of *d*, the distance between the QDs location and the NW end.

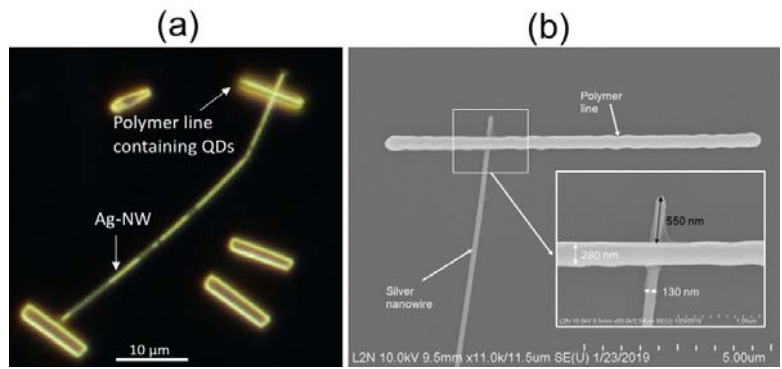

FIG. 3. Hybrid plasmonic nanostructures consisting of silver nanowire coupled with QD-containing polymer lines that were written by two-photon photopolymerization. (a) Dark-field white light optical image (b) SEM image. Inset: zoom-in at the crossing point between silver nanowire and polymer line.

QDs were used to launch SPs on Ag NWs. The optical set-up consists of a confocal microscope with a 405 nm excitation wavelength from a cw laser (Fig. 4 (a)). The incident laser beam was focused using a Nikon Plan 100x objective of N.A = 0.99. The long-pass filter at 450 nm allows one to collect the QDs emission only. A pinhole was used to select a specific detection area on the sample in order to collect the local photoluminescence. Both detection area and the focused laser beam have a diameter of about 250 nm, which is larger than the diameter of the Ag-NW and much smaller than the NW length.

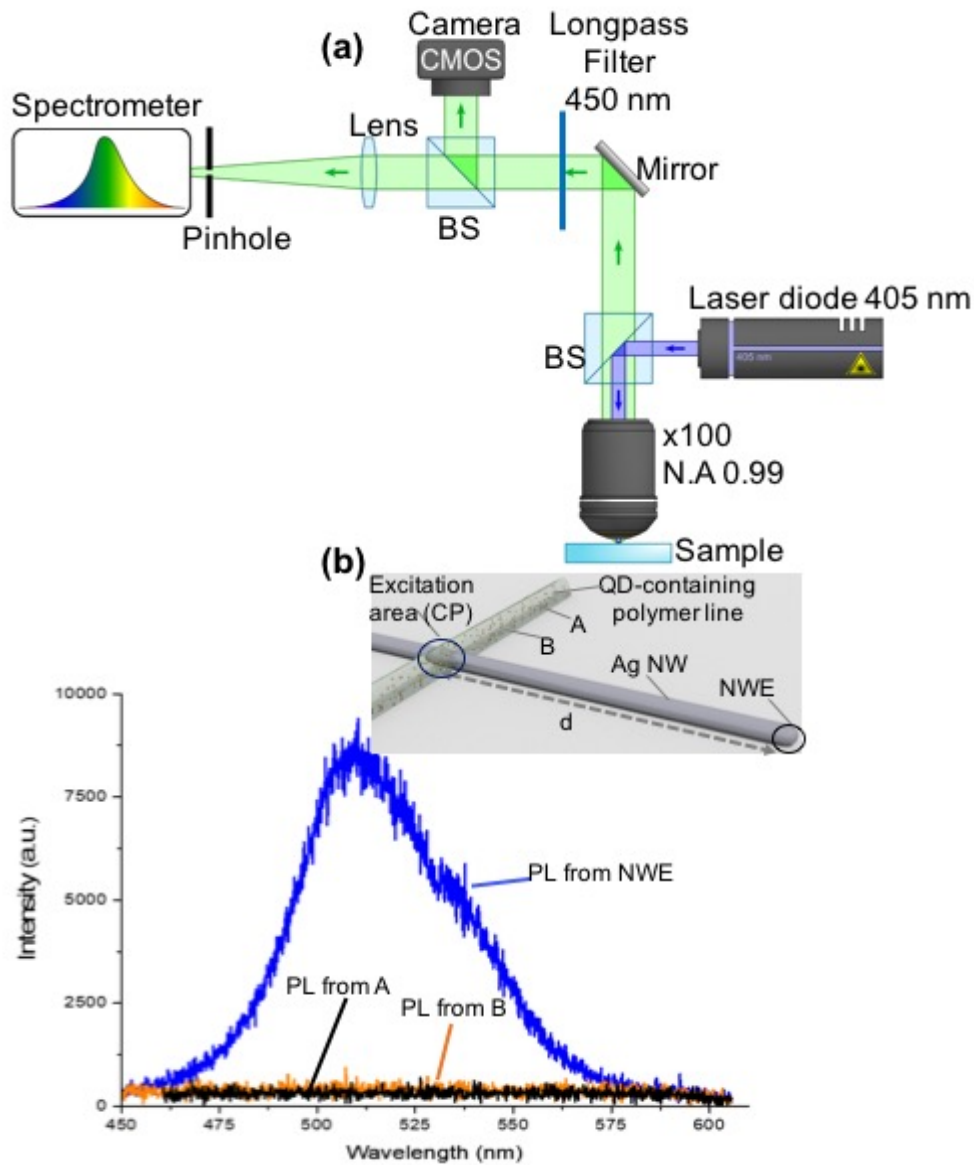

FIG. 4. Local photoluminescence (PL) from the hybrid plasmonic nanosytem (a) Optical set-up; (b) PL spectrum from different areas. Blue curve: from nanowire extremity (NWE). Black and orange curves: from the polymer line outside the nanowire (A and B locations). The excitation area (CP, 405 nm wavelength) is fixed. CP-NWE distance d is a controlled parameter.

Fig. 4 (b) shows different local PL spectra taken at different locations while the 405-nm wavelength excitation area was fixed and localized at the crossing point (CP) between the NW and the QD-containing polymer line. CP and the NW end (NWE) are separated from each other by a controlled distance *d*. Measurable PL spectra were obtained only at NWE (blue spectrum on Fig. 4 (b)) and at CP. In particular, no PL signature was observed at the polymer line (orange and black spectra on Fig. 4 (b)) and at the NW body, between CP and NWE.

The fact that PL was observed at the NW end suggests an energy transfer between excited QDs (at CP) and NW end *via* surface plasmons propagation: QDs were excited in close proximity to a Ag-NW, the QDs near-field emission was transferred into guided propagating NW surface plasmons that were scattered at the NW end. While NW surface plasmons were directly excited in the green in Fig. 2 (b,c), green surface plasmons were excited in an inelastic way, *via* QDs emission, in Fig. 4 (b). Compared with metallic nanoparticles, the Ag-NW enables the propagation of SPs in a well-defined direction along the NW,[12] making possible long-distance energy transfer between the nanoemitter and a specific NW point of interest. The possibility to place the nanoemitters in a controlled way is a clear asset as it will be illustrated in Fig. 5.

Energy transfer between QDs and metallic nanoparticle is governed by dipole-dipole interactions, and the transfer rate between CP and NWE is determined by both SP excitation efficiency (EE) at CP and surface plasmon propagation decay. At CP, orientation and position of QD dipoles is supposed to be random. EE depends on three main parameters: QD dipole orientations (since surface plasmons are longitudinal waves, mainly dipoles parallel to the NW are expected to excite the plasmon mode), their distance from the NW, and the spectral overlap between the plasmon modes and the QDs emission[24]. It is assumed that EE is constant whatever the position of CP (i.e. whatever *d*).

A parametric study of the plasmon propagation was performed. Several hybrid structures were made to study the influence of *d*, the distance between CP and NW end (see Fig. 4 (b)). In other words, for the first time, we controlled, by 2-PP, the position on the metal nanowire of the nanoemitters-containing site that acts as a plasmon launching site.

Fig. 5 shows the results of the study. Measured PL intensity detected at NWE is plotted as a function of *d* (black squares). Each point is the mean value of ten successive measurements on the same structure. It should be pointed out that *d* is actually the distance between NWE and the center of the polymer line whose width is about 280 nm (see Fig. 3 (b)). For each point, PL intensity at NWE was normalized by the PL intensity directly detected right at CP. The PL intensity at NWE exponentially decreases as *d* increases, which is the signature of a propagating surface plasmon whose intensity can be described as follows:[28]

$$I = I_0 e^{-\alpha x} = I_0 e^{-\frac{x}{L_{SPP}}} \quad (1)$$

With $\alpha$ the attenuation coefficient and $L_{SPP}$ the plasmon propagation length. In our case, the observable corresponds to *x=d*. It should be pointed out that $I_o$ depends on at least four parameters: incident exciting intensity, QD absorption rate and quantum yield (QY =0.6) and excitation efficiency

EE (defined above). EE, QY and absorption rate are supposed to be constant and measured intensity was normalized by incident intensity. $I_o$ is thus considered as constant for any *d* value.

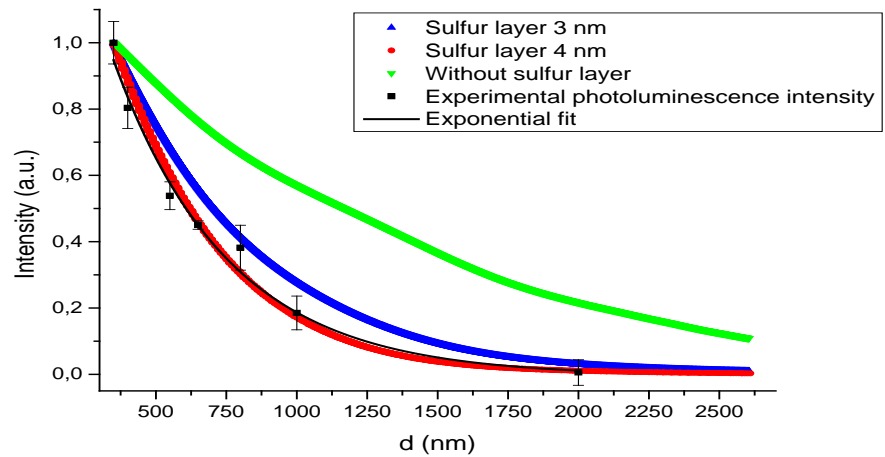

FIG. 5. PL Intensity at the nanowire extremity. Black dots: PL intensity measured at NWE as a function of the distance d between NWE and the polymer line. Black curve: exponential fit. Green curve: simulated intensity along a Ag-NW without sulfurization layer. Blue and red curves: simulated intensity along a Ag-NW with sulfurization layer (3 nm and 4 nm thickness respectively).

Fig. 5 shows a clear dependence on *d*, showing that it is possible to control the PL intensity at NWE by controlling the position of the launching site.

Due to the SP propagation loss, the emission intensity decays exponentially with respect to the propagation distance as shown by equation (1). By fitting the PL intensity using the exponential form (black curve in Fig. 5), the SPs propagation length can be quantified. In our study, the SP propagation length for excitation at 510 nm by QDs in polymer was found to be around 407 nm, which is surprisingly lower than expected. Indeed, typical SP propagation length on silver can reach a 1.1 µm value for green light.[29] Spatial decay of Ag-NW SP launched by a dipole was calculated by Finite Difference Time Domain (FDTD) method (see supplementary material). The resulting exponential decay shown in Fig. 5 (green curve) confirms a propagation length > 1 µm. This significant difference is interpreted as follows. SPs propagation length can strongly decrease due to either surface roughness or formation of silver sulfide at Ag-NW surface.[30,31] Effect of surface roughness can be ruled out for chemically synthesized Ag-NWs. However, the HCl acid used for the rinsing procedure probably removed the protective layer around the nanowires, making possible NWs sulfurization. We ran several Finite FDTD simulations with different sulfur layer thicknesses (see supplementary material). Each simulation was designed as follows: an Ag-NW is encased into a thin silver sulfide shell (n=3.1+0.65i[32]) and is placed onto a glass substrate. The sulfurized wire was then excited through a dipole placed close to its surface, which can thus get coupled to SPs. The field intensity along the wire was calculated. On Fig. 5, blue and red curves represent the calculated intensity decay in the presence of 3 nm and 4 nm thick layer of sulfur layers, respectively. The theoretical SP propagation length was accessed in both cases: 3 nm-thick sulfur layer leads to $L_{SPP}$ = 510 nm that drops to about 400 nm for the 4-nm sulfur thickness. SP propagation is thus strongly affected by the presence of this sulfur layer. We can deduce that the Ag-NWs have endured sulfurization and that the Sulfur layer is about 4 nm. This result implies that, in case Ag-NWs are used for nanophotonics applications, control of the atmosphere surrounding them is critical. Alternatively, Ag-NWs could be protected by a thin dielectric layer of controlled thickness.

In conclusion, Ag-NW supporting propagating SPs is a fundamental building block for plasmonic integrated circuits. NW can get coupled with light emitting QDs, providing a promising integrated platform for optical information propagation and processing. Our work shows that it is possible to place QDs on Ag-NW at controlled sites, by 2-photon photo polymerization of a designed QD-containing photosensitive formulation. First, we have shown experimentally that the QD emission can excite Ag-NW surface plasmons from a well-defined launching site. The launched SPs propagate along the NW axis. By scattering, light is emitted from the NW extremity. 2-PP allowed one to carry out a parametric study of the distance between the integrated QDs and the Ag-NW end. With this study, we showed that light intensity at the NW extremity can be controlled by controlling the position of the launching site, through SPs propagation length that could be precisely deduced.

Perspectives are numerous. In particular, much thinner polymer lines[21] will be used to identify launcher position in a more precise way and to integrate single QDs by controlling both line width and QD concentration within the formulation. In addition, this new approach is very promising to produce efficient acceptor-donor hybrid nano-systems by using different kinds of QD-containing photopolymerizable formulation. [20,21]

See supplementary material for a detailed description of FDTD calculation

Experiments were carried out within the Nanomat platform (www.nanomat.eu). This project is financially supported by the ANR (French research agency) and the NRF (National Research Foundation, Singapore) through the international ACTIVE-NANOPHOT (alias "MUHLYN") funded project (ANR-15-CE24-0036-01).